\documentclass[aps,showpacs,amsmath,amssymb,pra,superscriptaddress,twocolumn,floatfix,10pt]{revtex4-1}

\usepackage{graphicx}
\usepackage{hyperref}
\usepackage{braket}

\allowdisplaybreaks[1]

\begin{document}

\title{ ${\mathcal{PT}}$ symmetry breaking in waveguide arrays with competing loss/gain pairs}

\author{P.~A. Kalozoumis}
\affiliation{Department of Physics, University of Athens, GR-15771 Athens, Greece}

\author{C.~V. Morfonios}
\affiliation{Zentrum f\"ur Optische Quantentechnologien, Universit\"{a}t Hamburg, Luruper Chaussee 149, 22761 Hamburg, Germany}

\author{F.~K. Diakonos}
\affiliation{Department of Physics, University of Athens, GR-15771 Athens, Greece}

\author{P. Schmelcher}
\affiliation{Zentrum f\"ur Optische Quantentechnologien, Universit\"{a}t Hamburg, Luruper Chaussee 149, 22761 Hamburg, Germany}
\affiliation{The Hamburg Centre for Ultrafast Imaging, Universit\"{a}t Hamburg, Luruper Chaussee 149, 22761 Hamburg, Germany}

\date{\today}

\begin{abstract}
We consider a periodic waveguide array whose unit cell consists of a $\mathcal{PT}$-symmetric quadrimer with two competing loss/gain parameter pairs which lead to qualitatively different symmetry-broken phases. 
It is shown that the transitions between the phases are described by a symmetry-adapted nonlocal current which maps the spectral properties to the spatially resolved field, for the lattice as well as for the isolated quadrimer.
Its site-average acts like a natural order parameter for the general class of one-dimensional $\mathcal{PT}$-symmetric Hamiltonians, vanishing in the unbroken phase and being nonzero in the broken phase. 
We investigate how the beam dynamics in the array is affected by the presence of competing loss/gain rates in the unit cell, showing that the enriched band structure yields the possibility to control the propagation length before divergence when the system resides in the broken $\mathcal{PT}$ phase.
\end{abstract}


\maketitle

\section{Introduction}\label{section1}

The concept of non-Hermitian systems with  $\mathcal{PT}$-symmetry has developed into a rapidly evolving research field in contemporary physics, extending from quantum mechanics~\cite{Bender2002} and field theory~\cite{Bender2004} to systems with topological states~\cite{Schomerus2013,BZhu2014,Poli2015}, optics~\cite{Ganainy2007,Makris2010, Musslimani2008a, Musslimani2008b,Peng2014,Peng2013} and acoustics~\cite{Zhu2014,Zhu2015,Fleury2015}. 
In the seminal work of Bender {\it et al.}~\cite{Bender1998} it was demonstrated that a class of non-Hermitian Hamiltonians can possess, for certain parametric ranges, entirely real eigenvalue spectra, indicating the unbroken $\mathcal{PT}$-symmetric phase where the Hamiltonian $\mathcal{H}$ and the $\mathcal{PT}$ operator share the same set of eigenvectors. 
On the other hand, by varying the parameter which determines the amount of loss and gain, a spontaneous symmetry breaking occurs such that $\mathcal{H}$ and  $\mathcal{PT}$, though still commuting, possess different eigenvectors with complex energy eigenvalues. 

The progress in the field of $\mathcal{PT}$-symmetric systems has since been enormous, to a wide extent impelled by the adaptation of the corresponding concepts in optical structures and the experimental realization of spontaneous $\mathcal{PT}$ symmetry breaking~\cite{Ruter2010,Guo2009}. 
The link between classical optics and quantum mechanics here stems from the effective description of light propagation in waveguide systems by the Schr\"odinger equation, with time represented by the waveguide axis within the paraxial approximation \cite{Eisenberg2000}.
In the realm of discrete optics, one-dimensional (1D) $\mathcal{PT}$-symmetric photonic waveguide systems have been associated with remarkable phenomena such as double refraction~\cite{Makris2010}, power oscillations~\cite{Ruter2010} and non-reciprocal diffraction~\cite{Makris2008}.
In the presence of nonlinearity, unidirectional~\cite{Ramezani2010} and asymmetric ~\cite{Ambroise2012,Yang2015} wave propagation have been observed. 
Such intriguing properties are present in a large class of oligomers~\cite{Ambroise2012,Li2011,Horne2013}, i.e. single  $\mathcal{PT}$ dimers~\cite{Kottos2010,Ramezani2010}, trimers~\cite{Li2013} and quadrimers~\cite{Konotop2012,Zezyulin2012,Gupta2014,Gupta2015,Phang2015}.
Moreover, in extended $\mathcal{PT}$-symmetric lattices the occurrence of Bloch oscillations \cite{Longhi2009} and universality in beam dynamics~\cite{Zheng2010} have been reported.
The addition of a second loss/gain rate has been discussed in Refs.~\cite{Zezyulin2012,Zezyulin2013} where the focus was on nonlinear modes in finite $\mathcal{PT}$-symmetric systems and in~\cite{Bendix2010} where a closed-form quadrimer was addressed.
An open perspective is the effect of competing pairs of loss/gain elements on the band structure of lattice systems and on the induced beam dynamics.

$\mathcal{PT}$-symmetric wave mechanics has also been applied to light scattering, owing to the isomorphy between the Helmholtz and stationary Schr\"odinger equations.
With refraction index landscapes which obeying $n(x)=n^{*}(-x)$, which corresponds to balanced gain and loss~\cite{Gonzalo2005}, such $\mathcal{PT}$-symmetric setups have been shown to feature scattering properties such as simultaneous coherent absorption and lasing~\cite{Chong2011,Longhi2009} and anisotropic transmission resonances~\cite{Ge2012}.
In contrast to bound or periodic $\mathcal{PT}$-symmetric systems, here the energy (frequency) of the incoming wave is real by definition, and the unbroken phase is indicated by the field itself being in a $\mathcal{PT}$ eigenstate.
Considering the plethora of different types of $\mathcal{PT}$-symmetric settings in general wave mechanics, continuous or discrete systems with closed, periodic, or scattering boundary conditions, the need for a simple yet universal quantity that pinpoints the broken and unbroken $\mathcal{PT}$ phases in all cases---thus going beyond the Hamiltonian spectrum---becomes clear.

In the present work we show that the spatial average of a discrete, $\mathcal{PT}$-adapted nonlocal current can take on this role in general 1D lattice systems.
Together with its continuous counterpart proposed recently for scattering systems~\cite{Kalozoumis2014b,Kalozoumis2014a,Kalozoumis2013b,Kalozoumis2015a,Kalozoumis2015b,Kalozoumis2013a}, this quantity provides a natural order parameter for $\mathcal{PT}$-symmetric 1D systems, derived directly from their eigenstates, which universally describes the transition between broken and unbroken phases.
We here demonstrate its applicability for a periodic $\mathcal{PT}$-symmetric waveguide array with multiple, qualitatively different regimes of nonhermitian evolution owing to the interplay of two competing pairs of loss/gain elements within a quadrimer constituting the unit cell.
Using a  Bloch mode analysis we associate the light propagation  properties in extended waveguide arrays to the enriched band structure landscape which allows for the control of the propagation distance of Gaussian wave packets prior to their divergence in the $\mathcal{PT}$ broken phase.

The paper is organized as follows: 
In Sec.~\ref{section2} we introduce the setup and the nonlocal currents which are used to obtain the corresponding phase diagram of the isolated quadrimer cell.
In Sec.~\ref{section3} we investigate the lattice band structure for varying cell parameters and show that the symmetry breaking information can be extracted from the nonlocal currents of Bloch states. 
We then demonstrate how the loss/gain pair competition may be used to control the dynamics of designed wave packets in the broken phase.
In Sec.~\ref{section4} we conclude our results.
  
\section{$\mathcal{PT}$ symmetry breaking and nonlocal currents for competing loss/gain pairs}\label{section2}

\begin{figure}[t!]
\includegraphics[width=.7\columnwidth]{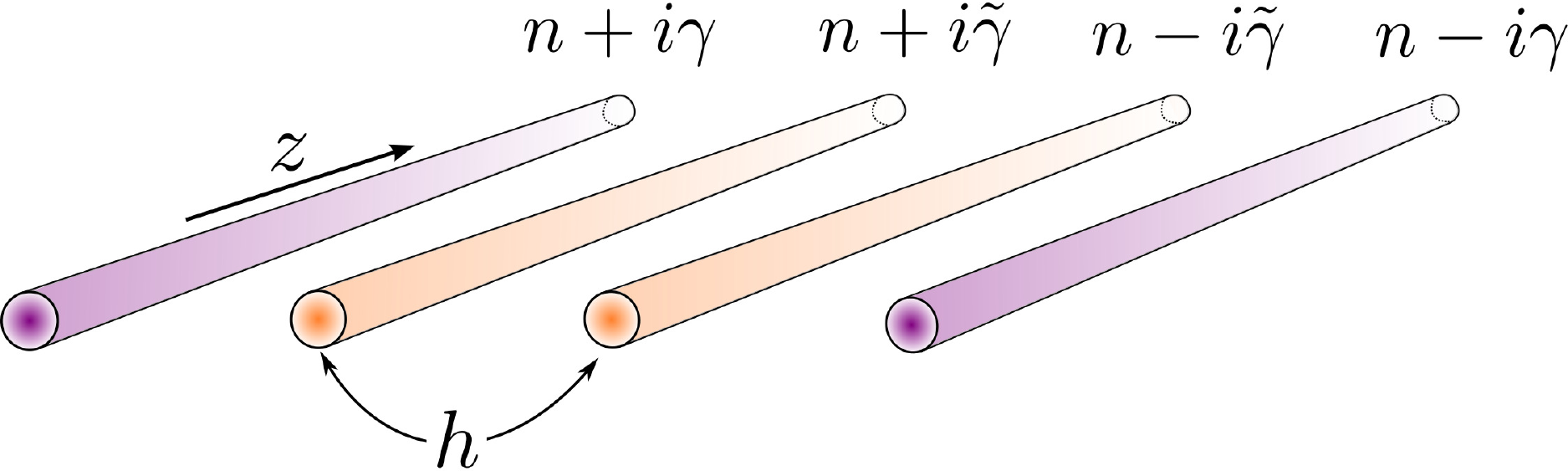}
\caption{\label{fig:sketch} (color online) 
Schematic of the setup, consisting of four waveguides with equal refraction indices $n$, uniform hoppings  $h$ and two competing loss/gain parameters $\gamma,~\tilde{\gamma}$.}  
\end{figure}

We start by analyzing a $\mathcal{PT}$-symmetric finite array of $N = 4$ photonic waveguides (quadrimer) as shown in Fig.\,\ref{fig:sketch}, which will later constitute the unit cell of a periodic array.
The quadrimer has on-site elements $\upsilon_{m}=n_{m} + i \gamma_{m}$ with common refraction index $n_{m}=n$ and \textit{two different} loss/gain rates $\gamma$ and $\tilde{\gamma}$ in $\mathcal{PT}$-symmetric arrangement: $\gamma_{1}=-\gamma_{4}=\gamma,~\gamma_{2}=-\gamma_{3}=\tilde{\gamma}$.
In the paraxial approximation \cite{Eisenberg2000}, the propagation of the light field along the $z$-axis is described by a set of four  ($m=1,2,3,4$) equations
\begin{equation}
\label{light_propag_1}  
 i\frac{d\psi_{m}}{dz}  =\upsilon_{m} \psi_{m} + h_{m,m+1} \psi_{m+1} + h_{m,m-1} \psi_{m-1},
\end{equation}
where $h_{m,m \pm 1}$ couples light from waveguide $m \pm 1$ to $m$. 
Considering common internal couplings $h_{m,m \pm 1} = h$ and $h_0 = h_{5} =0$, the state vector $\ket{\Psi} = (\psi_{1},\psi_{2},\psi_{3},\psi_{4})^\top$ of the isolated quadrimer is then governed by the Schr\"odinger equation $i\frac{d}{dz} | \Psi \rangle = {\mathcal{H}} | \Psi \rangle$ with Hamiltonian matrix
\begin{equation}
\label{hamiltonian} 
{\mathcal{H}} =  \left( \begin{array}{cccc}
\upsilon & h & 0 & 0 \\
h & \tilde{\upsilon} & h  & 0 \\
0 & h & \tilde{\upsilon}^{*} & h \\
0 & 0 & h  & \upsilon^{*}    \end{array} \right),
 \end{equation}
where ${\upsilon} = n + i{\gamma}$, $\tilde{\upsilon} = n + i\tilde{\gamma}$. 
If $| \Psi \rangle = | \Psi^{(j)} \rangle$ is a (right) eigenvector of $\mathcal{H}$ with eigenvalue $\varepsilon_{j}$, $\psi^{(j)}_{m}=\alpha^{(j)}_{m}e^{i \varepsilon_{j} z}$, then the close-coupling equations (\ref{light_propag_1}) become
\begin{align}
\varepsilon_{j} \alpha^{(j)}_{m} &=\upsilon_{m} \alpha^{(j)}_{m} +h\left(\alpha^{(j)}_{m+1}+\alpha^{(j)}_{m-1}\right). \label{static_Helm1}
\end{align}
Throughout the paper, we measure lengths, field amplitudes and ${\mathcal{H}}$-elements in units of $a^0$ (lattice constant), $\alpha^0$ (amplitude unit) and $\epsilon^0$ (${\mathcal{H}}$-element unit), respectively, which are in turn set to unity, $a^0 = \alpha^0 = \epsilon^0 = 1$.

The aim is now to extract the information on whether this state is $\mathcal{PT}$ broken or unbroken from mode amplitudes $\alpha^{(j)}_{m}$ alone, in terms of a single quantity.
To do so, we combine Eq.\,(\ref{static_Helm1}) with its complex conjugate at the $\mathcal{PT}$-related site $\bar{m}=N+1-m = 5 -m$ to construct the difference
\begin{equation} \label{q_with_e}
{Q}^{(j)}_{m}-{Q}^{(j)}_{m-1}=\left(\varepsilon_{j}-\varepsilon_{j}^{*}\right)\alpha^{(j)}_{m} \alpha^{*(j)}_{\bar{m}},
\end{equation}
where the $\mathcal{PT}$ symmetry $\upsilon_{m}=\upsilon_{\bar{m}}^{*}$ has been taken into account, with the \textit{nonlocal current} of mode $\alpha^{(j)}$ defined here as
\begin{equation} \label{q1} 
{Q}^{(j)}_{m} = h \left(\alpha^{(j)}_{m+1} \alpha^{*(j)}_{\bar{m}}-\alpha^{*(j)}_{\bar{m}-1} \alpha^{(j)}_{m}\right).
\end{equation}
If the eigenvalue $\varepsilon_{j}$ is complex, then ${Q}^{(j)}_{m}$ varies along the array, but for $\varepsilon_{j}=\varepsilon_{j}^{*}$ we have from Eq.\,(\ref{q_with_e}) that ${Q}^{(j)}_{m} = {Q}^{(j)}_{m-1}$, and so the nonlocal current is \textit{spatially constant} for real $\varepsilon_{j}$, and further vanishes, ${Q}^{(j)}_{m} = 0$, since $| \Psi^{(j)} \rangle$ is then a $\mathcal{PT}$ eigenstate.
Hence, we can use the site-averaged current $\braket{{Q}^{(j)}} = \sum_m {Q}^{(j)}_m/N$ as a single quantity which distinguishes the unbroken phase with $\braket{{Q}^{(j)}} = 0$ from the broken one with $\braket{{Q}^{(j)}} \neq 0$.
This extends the use of a symmetry-adapted nonlocal current as an order parameter for phase transitions in $\mathcal{PT}$-symmetric systems from scattering in continuous setups \cite{Kalozoumis2014b} to eigenmodes in discrete setups (note that no site average is needed for scattering where $\varepsilon$ is a real input parameter). 
A generic account on nonlocal currents in discrete models with symmetry domains will be given elsewhere~\cite{Morfonios2015}. 
Here we employ them to describe the spontaneous symmetry breaking in globally $\mathcal{PT}$-symmetric finite and periodic arrays.

\begin{figure}[t!]
\includegraphics[width=.7\columnwidth]{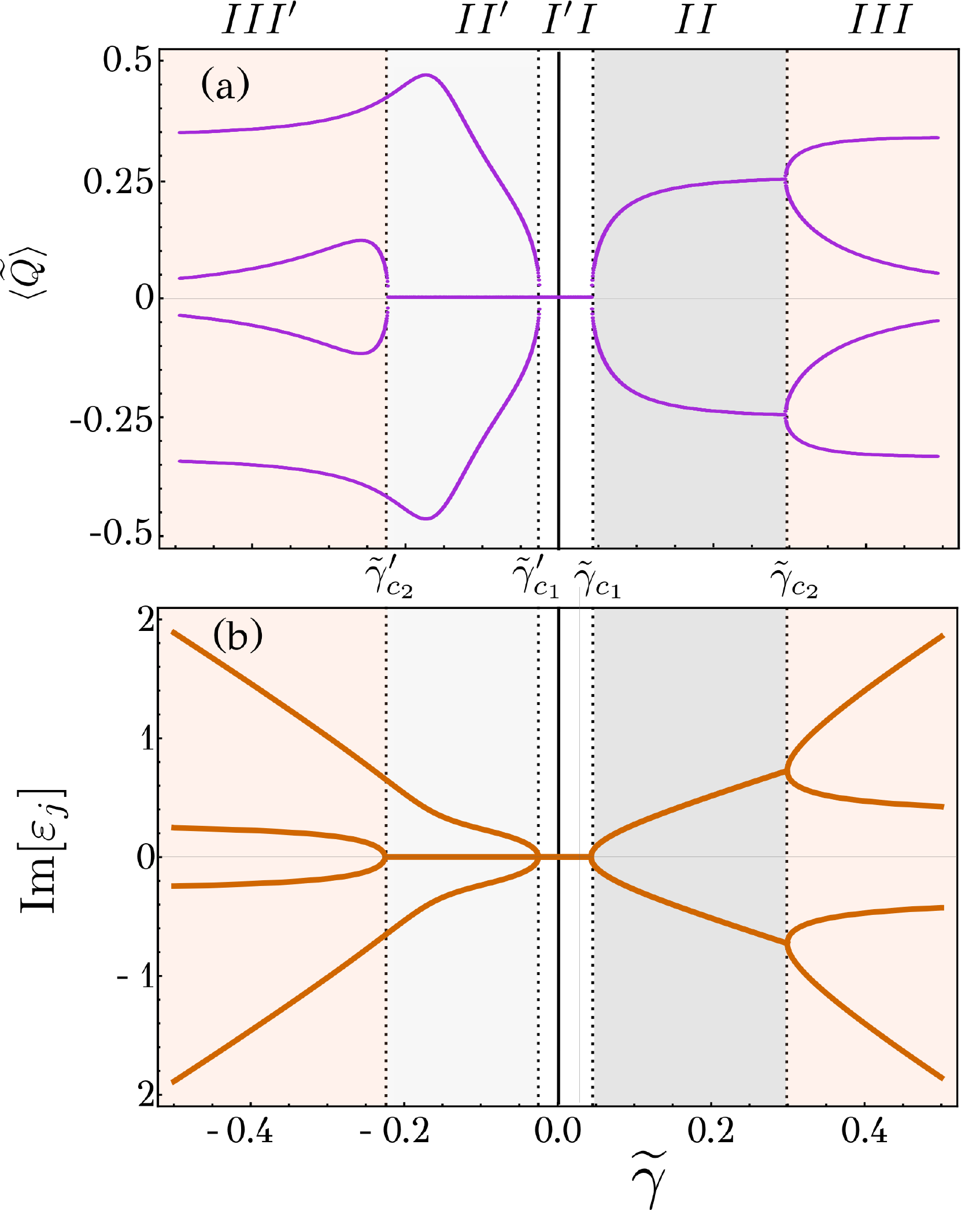}
\caption{\label{figq} (color online) (a) Phase diagram of the setup of Fig.\,\ref{fig:sketch} defined by the site-averaged current $\braket{{Q}} $ as a function of the loss/gain parameter $\tilde{\gamma}$, for $ \gamma = 0.08,~h = 0.1$ and $n = 0.1$. Dotted vertical lines indicate transition points $\tilde{\gamma}_{c_{i}}$ ($\tilde{\gamma}^{\prime}_{c_{i}}$), $i=1,2$, between different $\mathcal{PT}$ phases denoted by regions $I,II,III$ ($I',II',III'$) for $\tilde{\gamma}>0$ ($\tilde{\gamma}<0$). The region shadings highlight their qualitative difference: real ($I,I'$), complex ($II,II'$, with distinct partial degeneracy) and imaginary ($III,III'$) spectrum. (b) Imaginary part of the eigenvalue spectrum (in units of the hopping $h$) of the setup of Fig.\,\ref{fig:sketch} as a function of the loss/gain parameter $\tilde{\gamma}$. }
\end{figure}

Figure~\ref{figq} shows $\langle {Q}^{(j)} \rangle$ for the four eigenmodes of the quadrimer together with the imaginary part of its eigenenergies (given in App.\,\ref{sec:app_eigenvalues}) as a function of $\tilde{\gamma}$  (note that all $\braket{{Q}^{(j)}}$ are real for any length $N$ of the $\mathcal{PT}$-symmetric array). 
Variation of $\tilde{\gamma}$ in the presence of the additional loss/gain rate $\gamma$ allows for a variety of transitions between broken/unbroken or between different broken $\mathcal{PT}$ phases at multiple transition points, generally not symmetric around $\tilde{\gamma}=0$. 
For example, region $II'$ is characterized by two real $\varepsilon_j$ and a complex conjugate pair, indicating wave propagation which is qualitatively different from region $II$ with two conjugate eigenvalue pairs. 
The qualitatively different propagation properties in regions $I$ and $II$ are discussed in App.\,\ref{sec:app_single_excitation}, with the behavior around the transition at the exceptional point $\tilde{\gamma}_{c_{1}}$ highlighted in App.\,\ref{sec:app_EP}.
Regions $III$ and $III'$ both have two distinct conjugate pairs of imaginary eigenvalues, though arising from different types of transitions from $II$ and $II'$, with a double splitting of degenerate imaginary pairs (at $\tilde{\gamma}_{c_{2}}$) as opposed to a splitting and two branch continuations (at $\tilde{\gamma}'_{c_{2}}$).

As is seen, all four $\braket{Q^{(j)}}$ vanish in the $\mathcal{PT}$-unbroken phase with $\text{Im}[\varepsilon_{j}] = 0$ (regions $I',~I$), while following the profile of the $\text{Im}[\varepsilon_{j}] \neq 0$ in the $\mathcal{PT}$-broken phases.
The nonlocal current is thus a quantity that exposes the nonhermitian part of the discrete spectrum although being constructed solely from the spatially resolved wavefunction with no direct reference to the onsite potential or the mode eigenvalues.
It equally characterizes, however, $\mathcal{PT}$ phase transitions for scattering states with real energy, as shown in Ref.\,\cite{Kalozoumis2014b}, thus unifying the description of $\mathcal{PT}$-symmetric bound and unbound systems (instead of employing either the Hamiltonian or $S$-matrix eigenvalues, respectively).

Let us now elaborate further on the information contained in the quantities $\braket{Q^{(j)}}$.
Specifically, they can be used to express symmetry \textit{remnants} in the nondegenerate phases of broken $\mathcal{PT}$ symmetry, where the individual $\braket{Q^{(j)}}$ are nonzero.
Due to the commutation $[\mathcal{PT},\mathcal{H}] = 0$, in the $\mathcal{PT}$-broken phases with complex $\varepsilon_{j}$ (such as regions $II,II',III,III'$ of Fig.\,\ref{figq}) we have that eigenstate pairs $\ket{\Psi^{(j)}},\ket{\Psi^{(j')}}$ with complex conjugate eigenenergies $\varepsilon_{j} = \varepsilon_{j'}^*$ are related by $\ket{\Psi^{(j)}} = \mathcal{PT}\ket{\Psi^{(j')}}$.
This remnant of $\mathcal{PT}$ symmetry in the broken phase is encoded in the nonlocal currents:
Whereas both ${Q}^{(j)}_m,{Q}^{(j')}_m$ of a $\mathcal{PT}$-related pair are nonzero and vary spatially, their sum always vanishes,
\begin{equation} \label{q_diff_modes}
Q^{(j)}_m + Q^{(j')}_m = 0 ~~~ \forall \,m,
\end{equation}
which is due to the $\mathcal{PT}$-antisymmetry of $Q^{(j)}_m$ itself.
At the level of the average nonlocal currents we have $\braket{Q^{(j)}} + \braket{Q^{(j')}} = 0$ as a single parameter describing the symmetry remnant in each subspace of $\mathcal{PT}$-related mode pairs.
Equivalently, we can express the symmetry remnant of the system by the vanishing sum $\sum_j \braket{Q^{(j)}}$ over all modes, signalizing the invariance of the complete eigenspace of the Hamiltonian under $\mathcal{PT}$ operation (up to a phase for $\mathcal{PT}$-unbroken modes).
This further demonstrates the capability of the $\braket{Q^{(j)}}$ to capture features related to the $\mathcal{PT}$ symmetry breaking and indicates that they can serve as a generic tool for a systematic approach to symmetry breaking in 1D systems.

At this point, let us comment on some similarities
which can be identified between the transition from the $\mathcal{PT}$ unbroken to the $\mathcal{PT}$ broken phase and the thermodynamic phase transitions in the mean field approach (e.g. Landau theory~\cite{Goldenfeld}) where critical fluctuations are absent: (i) In both cases there exists a quantity which is zero (nonzero) in the phase of the unbroken (broken) symmetry, this being the order parameter $\braket{O}$ in thermodynamics (e.g. the mean magnetization for the Ising case), while in $\mathcal{PT}$ symmetry breaking this role is assigned to $\braket{{Q}}$. (ii) In the vicinity of the transition from unbroken to broken phases  both $\braket{O}$ and $\braket{{Q}}$ follow a power law with exponent $1/2$ (in the present case the explicit form is $\braket{{Q}} \sim \vert \widetilde{\gamma}-\widetilde{\gamma}_{c} \vert^{1/2}$).
The same occurs in a multitude of $\mathcal{PT}$ transition scenarios for, e.\,g., single dimers or quadrimers with $\gamma=\widetilde{\gamma}$ or with $\gamma=-\widetilde{\gamma}$.
Such common transition features arise for spontaneous symmetry breaking upon varying a suitable parameter (loss/gain rates in $\mathcal{PT}$-symmetric systems and temperature in thermodynamics), as a result of the absence of interactions and criticality in the mean field approximation.

Before proceeding to the extended lattice in the next section, we briefly address the relation of the so called quasipower $P_{Q} = \sum_{m=1}^{N} \psi_{m}\psi^{*}_{\bar{m}}$ to the nonlocal currents \cite{Morfonios2015}, here based on Eq.\,(\ref{q_with_e}).
The quasipower has been shown~\cite{Bagchi2001} to be conserved along $z$ in $\mathcal{PT}$-symmetric systems for any propagating excitation and any choice of loss/gain parameter values, in contrast to the usual power $P = \sum_{m=1}^{N}|\psi_{m}|^2$ which generally oscillates in the $\mathcal{PT}$ symmetric phase or grows in the  $\mathcal{PT}$ broken phase~\cite{Makris2008}.
To study the behavior of the quasipower of a single mode $j$, its nonlocal current $Q^{(j)}$ can be employed: 
Summing Eq.\,(\ref{q_with_e}) over $m$ yields
\begin{equation}
 \label{quasiP_zero_proof} {Q}^{(j)}_{N}-{Q}^{(j)}_{0} = 2i\,\textrm{Im}[\epsilon_{j}] \,P_{Q}^{(j)},
\end{equation}
where $P_{Q}^{(j)}=\sum_{m=1}^{N}\alpha_{m}^{(j)}\alpha^{*{(j)}}_{N+1-m}$ (with $N = 4$ for the quadrimer). 
Now, since $\alpha_{0}=\alpha_{N+1}=0$ due to the boundary conditions, it follows that  ${Q}_{0} = {Q}_{N} = 0$, showing that the quasipower vanishes for any eigenmode in the $\mathcal{PT}$ broken phases of a finite array. 

\section{Band structure and beam dynamics}\label{section3}

We now turn to the periodic lattice with the quadrimer above as the unit cell.
The aim is to investigate how the competing loss/gain pairs affect the band structure and accordingly the beam dynamics in large waveguide arrays, as well as the existence of nonlocal currents which describe the occurrence of $\mathcal{PT}$ eigenstates.
The index $m$ now enumerates the repeated quadrimers coupled by intercell hopping $\eta$, and wave fields $A_{m}(z),~B_{m}(z),~C_{m}(z),~D_{m}(z)$ are assigned (from the left) to the waveguides of the $m$-th quadrimer. 
The coupled mode equations describing the evolution of the fields along the $z$-axis are then
\begin{equation}
\begin{aligned}
\label{band1} i \frac{dA_{m}(z)}{dz} &=\upsilon A_{m}(z) + \eta D_{m-1}(z) + h B_{m}(z)  \\
 i \frac{dB_{m}(z)}{dz}&=\tilde{\upsilon} B_{m}(z) + h A_{m}(z) + h C_{m}(z)  \\
 i \frac{dC_{m}(z)}{dz}&=\tilde{\upsilon}^{*} C_{m}(z) + h B_{m}(z) + h D_{m}(z) \\
i \frac{dD_{m}(z)}{dz}&=\upsilon^{*} D_{m}(z) + h C_{m}(z) +\eta A_{m+1}(z).
\end{aligned}
\end{equation}
To obtain the band structure of the system we use the Fourier transform of the fields in the first Brillouin zone (BZ), $A_{m}(z)=\frac{1}{2\pi} \int_{-\pi/L}^{\pi/L}  \tilde{A}_{k}e^{i k L m} dk$ etc. with $L = 4$ being the lattice period, so that Eqs.~(\ref{band1}) yield the evolution of $\ket{\tilde{\Psi}_k} = (\tilde{A}_{k},\tilde{B}_{k},\tilde{C}_{k},\tilde{D}_{k})^\top$ according to $i \frac{d}{dz}\vert \ket{\tilde{\Psi}_k}= {\mathcal{H}}_{k} \vert \tilde{\Psi}_{k} \rangle$ with Bloch Hamiltonian
\begin{small}
\begin{equation}
\label{band2}  {\mathcal{H}}_{k} =  \left( \begin{array}{cccc}
\upsilon & h & 0 & \eta e^{-ikL} \\
h & \tilde{\upsilon} & h  & 0 \\
0 & h & \tilde{\upsilon}^{*} & h \\
\eta e^{ikL} & 0 & h  & \upsilon^{*}   \end{array} \right).
\end{equation}
\end{small} 

\begin{figure}[t!]
\includegraphics[width=\linewidth]{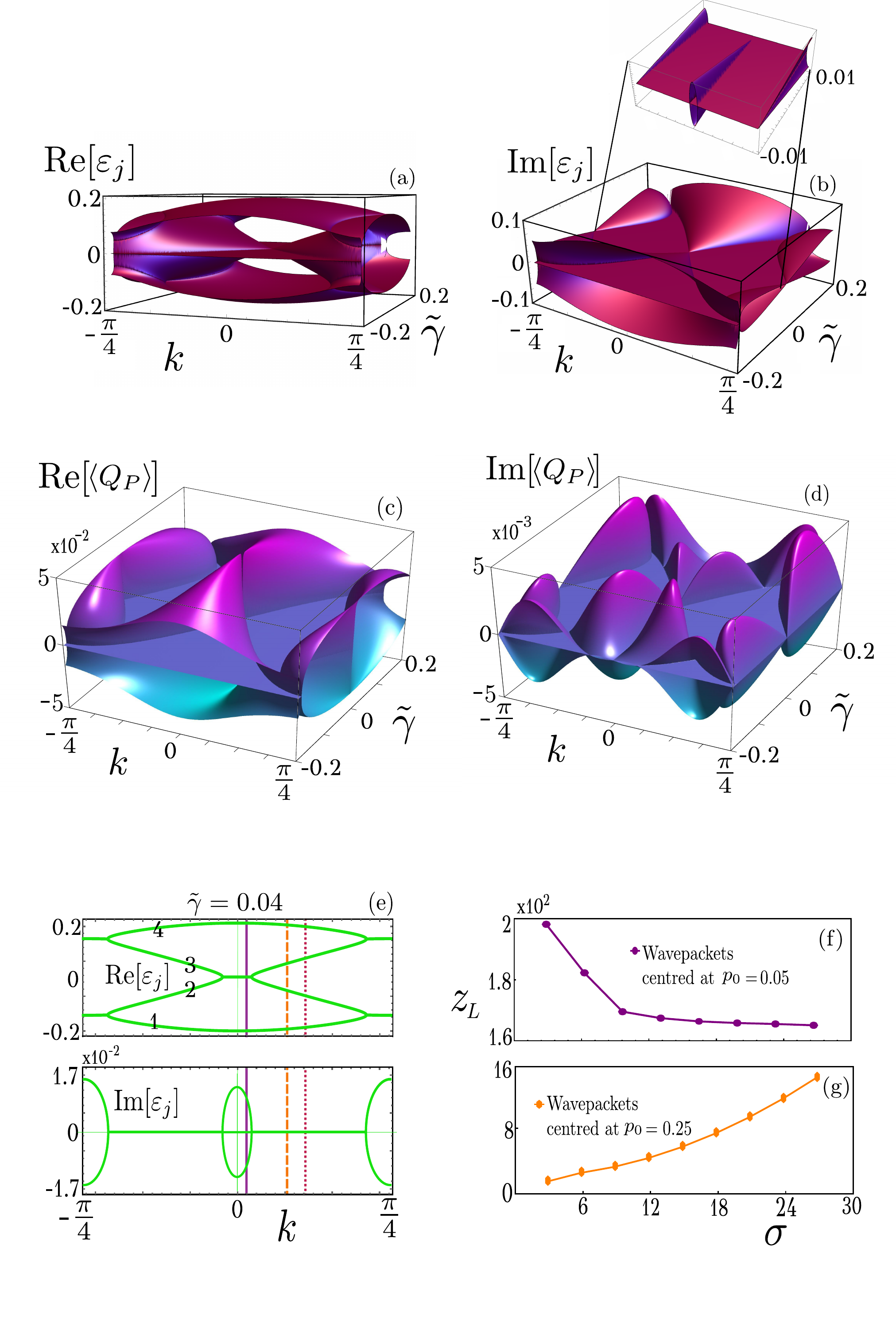}
\caption{\label{fig:bands} 
(color online) (a) Real and (b) imaginary parts of the dispersion relation $\varepsilon_{j}(k)$ of the periodic waveguide array with a quadrimer unit cell as a function of the loss/gain parameter $\tilde{\gamma}$, for $\gamma=0.01$, $n=0.1$, and $h=\eta=0.1$. 
The inset shows the band structure around zero, elucidating the cases $\tilde{\gamma}=\gamma$ and $\tilde{\gamma}=-\gamma$.
(c) Re$[\braket{Q^{(j)}_{P}}]$ and (d) Im$[\braket{Q^{(j)}_{P}}]$ of Bloch modes $j=1,2,3,4$ for varying $\tilde{\gamma}$. 
(e) Band structure for $\tilde{\gamma}=0.04$ with enumerated bands.
(f,\,g) Propagation length $z_L$ up to power divergence for varying width $\sigma$ of a Gaussian wave packet launched at the center of an array of $N = 400$ waveguides with initial momentum (f) $p_0=0.05$ and (g) $p_0=0.25$, indicated by the solid and dashed vertical lines in (e), respectively.}
\end{figure}

The four eigenvalues $\varepsilon_{j}(k)$ ($j = 1,2,3,4$) of $ {\mathcal{H}}_{k}$ constitute the band structure of the system, whose real and imaginary parts are shown for varying $\tilde{\gamma}$ Fig.\,\ref{fig:bands} (a) and (b), respectively, with the competing loss/gain parameter $\gamma$ kept fixed.
As we see, the bands have imaginary branches for any $\tilde{\gamma} \neq 0$, indicating the general absence of a $\mathcal{PT}$-symmetric phase (that is, Im$[\varepsilon_{j}(k)]=0$ $\forall k$) for the chosen parameters (this is a general feature for $\eta = h$).
In particular, the interplay of $\gamma,~\tilde{\gamma}$ causes the coexistence of real and complex energy regions. 
The latter appear around the center and edges of the BZ with their size tuned by $\tilde{\gamma}$, which in turn determines the width of the $k$-regions with Im$[\varepsilon_{j}(k)]=0$. 
Note that for $\gamma=\tilde{\gamma}$ ($\gamma=-\tilde{\gamma}$) complex energy regions appear only at the edges (center) of the BZ (see inset in Fig.\,\ref{fig:bands}\,(b)), indicating the decisive role of the $\gamma,~\tilde{\gamma}$ coexistence on the shape of the bands. 

The enriched dispersion relation induced by the loss/gain parameter competition thus offers an additional degree of freedom for the manipulation of the regimes of stable propagation, as we will see below. 
Let us first, however, investigate whether the nonlocal currents for the periodic system $\braket{Q^{(j)}_{P}}$ do capture the $\mathcal{PT}$-symmetry breaking characteristics as they did for the isolated quadrimer above.
Taking the $Q_{m}^{(j)}$ expressions of Eq.\,(\ref{q1}) and employing the Bloch theorem, the evaluation of $\braket{Q^{(j)}_{P}}$ for the corresponding four Bloch modes ($j=1,2,3,4$) is straightforward. 
Figures~\ref{fig:bands} (c) and (d) show the Re$[\braket{Q^{(j)}_{P}}]$ and Im$[\braket{Q^{(j)}_{P}}]$, respectively, as a function of $k$ and $\tilde{\gamma}$.
Indeed, the characteristics which are related to the breaking of $\mathcal{PT}$ symmetry are accurately captured by $\braket{Q^{(j)}_{P}}$, since the $k$ regions where $\braket{Q^{(j)}_{P}}=0 $ (flat areas in Fig.\,\ref{fig:bands}\,(c),(d)), which indicate that the $\ket{\tilde{\Psi}_k}$ are $\mathcal{PT}$ eigenstates, coincide with the real regions of the corresponding bands $\varepsilon_{j}(k)$ (flat area in Fig.\,\ref{fig:bands}\,(b)).
This confirms the applicability of the nonlocal currents also in periodic lattice systems as state-derived indicators of $\mathcal{PT}$-broken/unbroken parts in the BZ.

The key advantage of the coexistence competing loss/gain parameters pairs, as indicated above, is the flexibility to engineer the band with desired regions of complex energies around the center and edges of the BZ. 
This suggests the possibility to design setups where suitably prepared beams will propagate substantial distances (in fact, beyond the experimentally realized waveguide length) without divergence even in the $\mathcal{PT}$ broken phase.
To elaborate on this concept via an exemplary case we focus on the band structure for $\tilde{\gamma}=0.04$, shown in Fig.\,\ref{fig:bands} (e), with complex band branches around the center ($|k| \lesssim 0.075$) and edges ($|k| \gtrsim 0.66$).
The distance $z_{L}$ where divergence of a launched wave packet sets in (which we define by the doubling of optical power, $P(z_L)\equiv 2P(0)$) depends on how well it is localized around its initial momentum $p_{0}$ and in turn whether $p_{0}$ lies on (or close to) band branches with Im$[\varepsilon_{j}] < 0$ causing exponential increase (recall that a mode $j$ evolves in $z$ as $e^{i \varepsilon_{j} z}$). 

Fig.\,\ref{fig:bands} (f), (g) shows the distance $z_{L}$ for a Gaussian wave packet $ A(n;z=0)=\left(\sqrt[4]{\pi} \sqrt{\sigma}\right)^{-1}e^{i p_0 n} e^{-i (n-n_{0})^{2}/2 \sigma^{2}}$ as a function of its width $\sigma$, for different initial momenta $p_{0}$.
The array is chosen large enough to enable a connection between the basic wave packet propagation properties and the features of the given band structure. 
For a small momentum $p_0=0.05$, the distance $z_{L}$ decreases with increasing $\sigma$ (Fig.\,\ref{fig:bands}\,(f)).
In contrast, for a moderate momentum $p_0=0.25$ the trend of $z_{L}$ is reversed, with the propagation distance now increasing with $\sigma$ (Fig.\,\ref{fig:bands}\,(g)). 
As we see, the wave packet can indeed travel a long distance $z_L$ without diverging (as in Fig.\,\ref{fig:bands}\,(f) for small $\sigma$), even though an unbroken $\mathcal{PT}$-symmetric phase (defined by $\text{Im}[\varepsilon_j(k)]=0~\forall j,k$) is overall absent in the band structure.
In other words, propagation properties which would be expected only in the $\mathcal{PT}$-symmetric phase of a setup may also be encountered as transient phenomena in the much larger parameter regions of $\mathcal{PT}$ broken phases. 
In the following, we analyze this behavior in terms of the Bloch state composition of a launched beam.

\begin{figure}[t!]
\includegraphics[width=.9\linewidth]{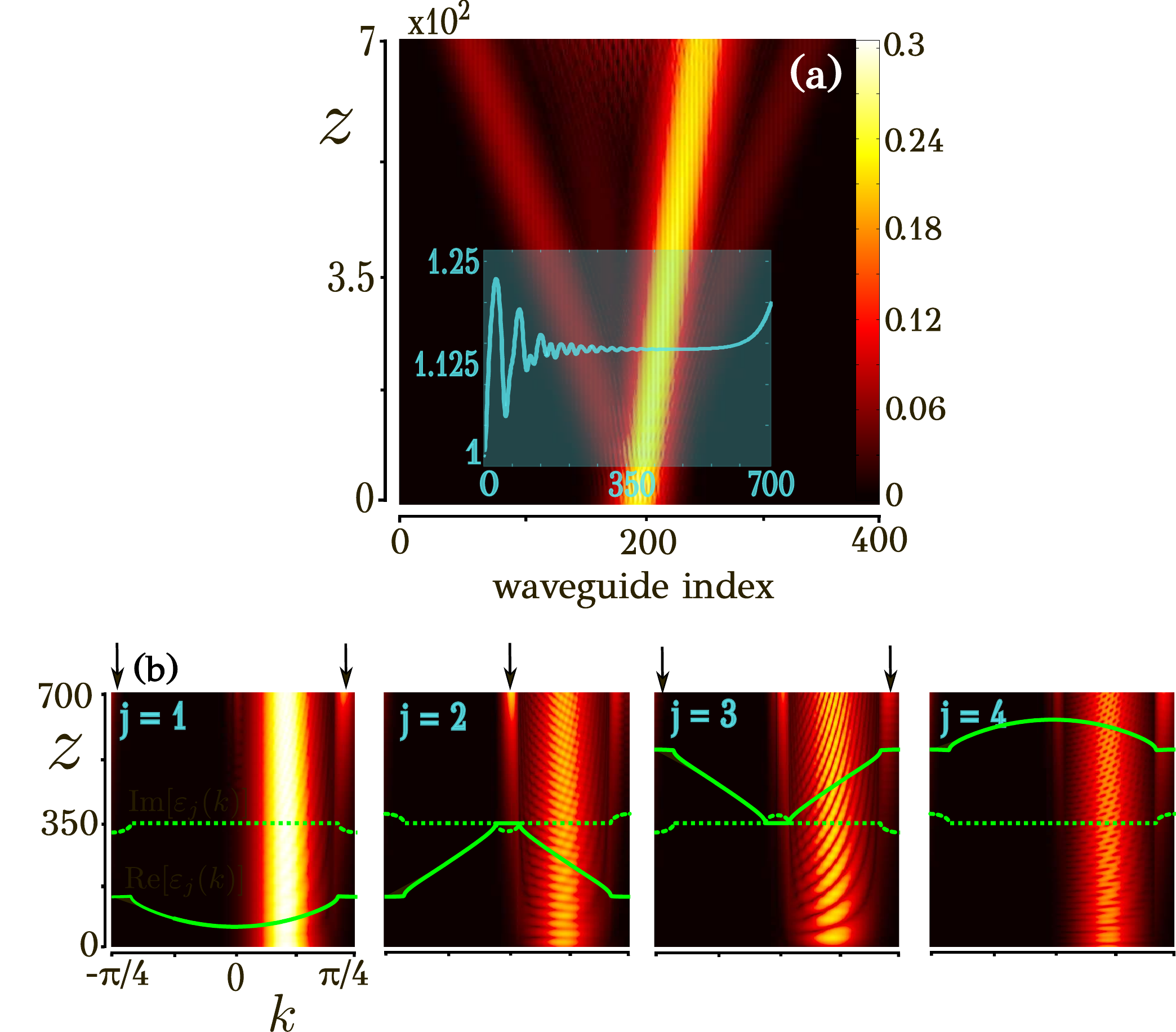}
\caption{\label{fig:propagation_lattice} 
(color online) (a) Propagation of a Gaussian wave packet of width $\sigma=13$ launched from the center of an array of $N=400$ waveguides with initial momentum $p_0=\pi/9$ and corresponding optical power evolution $P(z)$ (inset). (b) Expansion coefficients $|c_j(k)|^2$ in the right Bloch states of the corresponding periodic system, in each of the four bands $j=1,2,3,4$. Green dashed (solid) lines show the imaginary (real) part of each band multiplied by $1.5 \times 10^3$. Arrows indicate the gradual population of the regions with Im$[\varepsilon_{j}(k)]<0$ causing the exponential increase.}
\end{figure}

Since the associated Bloch momenta are not conserved in the finite periodic setup, the decomposition of a wave packet in $k$-space will evolve in time, thus eventually extending to other regions in the BZ.
If the wave packet is spatially broad and strongly localized in $k$-space, the dispersion (Fig.\ref{fig:bands} (e)) suggests that it will  diverge after a relatively long (short) propagation distance $z_L$ when it is initially localized in a $k$-region without (with) Im$[\varepsilon_{j}(k)]<0$.
To illustrate this mechanism in the beam dynamics, we show in Fig.\,\ref{fig:propagation_lattice}\,(a) the evolution of a relatively broad wave packet ($\sigma=13$) with initial momentum $p_{0}=\pi/9$ (indicated by the dotted vertical line in the band structure of Fig.\,\ref{fig:bands}\,(e)) up to the length of power divergence (as depicted in the inset) together with its Bloch state coefficients $|c_j(k)|^2$ in Fig.\,\ref{fig:propagation_lattice}\,(b).
The wave packet is split into four beams of different intensities, with slopes corresponding to the group velocities of (the real parts of) the dispersion regions populated in each of the four bands.
One main beam is more intense, which is reflected also in its $k$-space decomposition in the first band.
After some propagation distance ($z \gtrsim 400$) the light field starts populating Bloch momenta in the $k$-regions with complex band branches.
In particular, regions with Im$[\varepsilon_{1}(k)]<0$ (indicated by arrows) are increasingly populated with $z$, thus leading to exponential divergence of the light intensity in the corresponding $k$-space contribution.

The above analysis demonstrates that $\mathcal{PT}$-symmetric periodic lattices can be designed for quasistable propagation of appropriately launched wave packets even in the $\mathcal{PT}$-broken phase of the stationary dynamics, parametrically tunable in terms of different loss/gain parameters.

\section{Conclusions}\label{section4}

We have shown that spontaneous $\mathcal{PT}$ symmetry breaking in isolated $\mathcal{PT}$-symmetric discrete wave mechanical systems can be described by symmetry-adapted nonlocal currents $Q_m$ which are zero (nonzero) in the $\mathcal{PT}$ unbroken (broken) phase.
The spatial average $\braket{Q}$ then constitutes a single parameter which is constructed by the field amplitudes and unambiguously parametrizes the $\mathcal{PT}$-phase transition with values following the behavior of the imaginary part of the eigenvalue spectrum. 
With its continuous version describing the $\mathcal{PT}$-phase transition also in scattering systems \cite{Kalozoumis2014b} with real wave frequencies, $\braket{Q}$ is proposed as a natural ``order parameter'' for symmetry breaking in the generic class of $\mathcal{PT}$-symmetric 1D systems (discrete or continuous, bound or scattering).

The introduced concepts were applied to a photonic waveguide quadrimer with two competing pairs of loss/gain elements, employed as the unit cell of a periodic lattice.
The $\braket{Q}$ here further reveal remnants of the spontaneously broken symmetry in subspaces of eigenstates of the quadrimer, and the characterization of $\mathcal{PT}$ eigenstates by vanishing $\braket{Q}$ carries over to Bloch states of the lattice.
The additional competing loss/gain parameter $\tilde{\gamma}$ was shown to enrich the band structure, giving rise to additional $k$-space regions with complex energies which are absent for a single loss/gain rate and whose extent can be tuned by $\tilde{\gamma}$. 
Employing a Bloch mode analysis and guided by the complex band structure we finally demonstrated how a wave packet can be designed in order to control its propagation distance before starting to diverge. 
This offers the possibility to observe propagation properties in the parametrically extended $\mathcal{PT}$-broken phases which are typically accessible only in the $\mathcal{PT}$-unbroken phase.

\section{Acknowledgements}

We are thankful to A.\,V. Zampetaki for illuminating discussions.
P.A.K acknowledges financial support from IKY Fellowships of Excellence for Postdoctoral Research in Greece - Siemens Program.

\appendix

\section{Light propagation in the quadrimer} 
\label{sec:appendix}

In this appendix we briefly analyze the spectral and light propagation properties of the isolated quadrimer of Fig.\,\ref{fig:sketch}.
First the quadrimer eigenvalues are provided and subsequently the qualitative difference of light propagation in the $\mathcal{PT}$-broken and unbroken phases is highlighted with a focus on the transition between them. 

\subsection{Quadrimer eigenvalues}
\label{sec:app_eigenvalues}

Diagonalization of the matrix ${\mathcal{H}}$ yields the following four energy eigenvalues:
\begin{equation}
\label{eigenvalues}  
\varepsilon_{j}=\pm \frac{1}{\sqrt{2}}  \left( \sqrt{ a \pm \sqrt{ b } }+ n \right) ~~;~~j=1,2,3,4,  \\
\end{equation}
where
\begin{align*}
a &= -(\gamma^{2}+\tilde{\gamma}^{2}-3h^{2}), \\
b &= \left(\gamma^{2}-\tilde{\gamma}^{2} \right)^{2} - 2\left(\gamma+\tilde{\gamma}\right)\left(\gamma+3\tilde{\gamma} \right)h^{2} +5h^{4}.
\end{align*}
The eigenvalues are drastically affected by the combination of $\gamma$ and $\tilde{\gamma}$, with the variation of $\text{Im}[\varepsilon_j]$ with $\tilde{\gamma}$ for fixed $\gamma$ displayed in Fig.\,\ref{figq}\,(b).

\subsection{Evolution of single waveguide excitation}
\label{sec:app_single_excitation}

\begin{figure}[t!]
\includegraphics[width=.95\columnwidth]{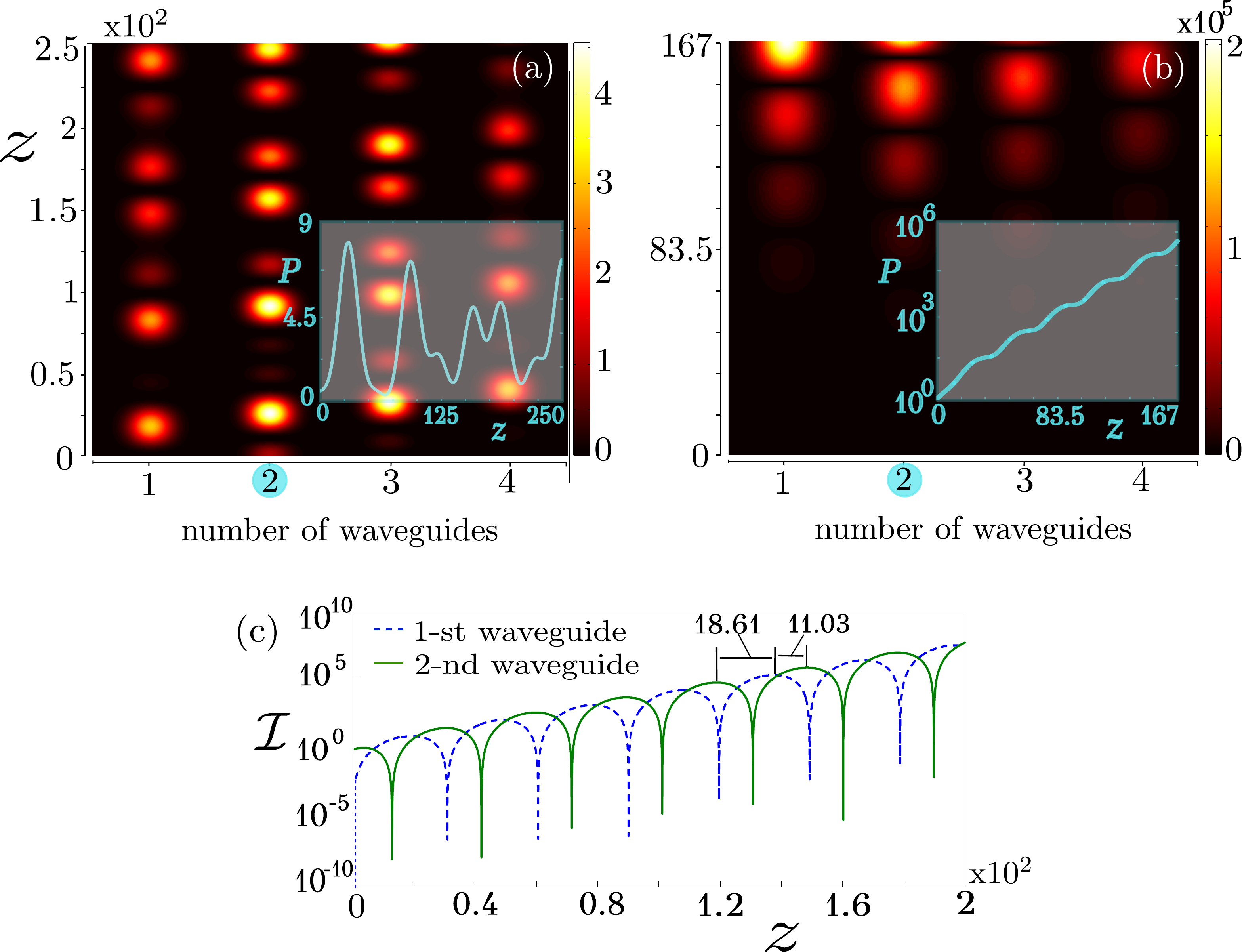}
\caption{\label{fig:quadrimer_propagation} (color online) Light propagation along the waveguide quadrimer for $h=0.1,~n=0.1,~\gamma=0.08,$ with light launched in the (marked) second waveguide. (a) $\mathcal{PT}$-unbroken phase with $\tilde{\gamma}=0.015$: Intensity $\mathcal{I}(x,z) = |\Psi(x,z)|^2$ smoothened along $x$ by convolution with a Gaussian function (color plot) and power $P$ (inset) as a function of propagation length $z$. (b) $\mathcal{PT}$ broken phase with $\tilde{\gamma}=0.06$: Intensity evolution up to the length $z=167$  where the sixth maximum in the first waveguide occurs, and the corresponding power (inset). (c) Intensity evolution in the first (blue, dashed lined) and second (green, solid line) waveguide for the parameter choice in (b): in each waveguide the maxima are equidistant with separation $2\pi/\text{Re}[{\epsilon}_2-{\epsilon}_1] \approx 29.64$, with the maxima of the first waveguide lagging by $18.61$ length units. 
}
\end{figure}

Since the eigenvectors of a non-Hermitian system are not orthonormal, we use the biorthogonal basis employing the left and the right eigenvectors of $\mathcal{H}$~\cite{Zheng2010} to compute the light propagation.
Figure~\ref{fig:quadrimer_propagation} illustrates the intensity evolution and the corresponding power $P$ (insets) following a unit excitation in the second waveguide.
In the $\mathcal{PT}$-unbroken parameter regions $I$ and $I'$ (Fig.\,\ref{fig:quadrimer_propagation}\,(a)) the light propagates without divergence and the power oscillates as expected~\cite{Kottos2010}. 
In contrast, the purely imaginary spectrum in the $\mathcal{PT}$-broken regions $III$ and $III'$ leads to a monotonous exponential increase of intensity and power with the propagation distance.
Of particular interest are the characteristics of the propagation in the $\mathcal{PT}$-broken region $II$.  
A key characteristic is here an exponentially growing light intensity with a superimposed oscillation, caused by the imaginary and real parts of the eigenvalues, respectively. 
This propagation pattern causes the intensity maximum to oscillate between the two gain waveguides along $z$, as shown in Fig.\,\ref{fig:quadrimer_propagation} (b). 
With the real part of the energy spectrum in this regime being doubly degenerate, the oscillations correspond to the single real energy difference yielding a period $2\pi/\text{Re}[{\epsilon}_2-{\epsilon}_1] \approx 29.64$.
As indicated in Fig.\,\ref{fig:quadrimer_propagation} (c), the oscillating pattern is also shifted by a constant length between the first and second waveguides.

\subsection{Propagation around exceptional point}
\label{sec:app_EP}

\begin{figure}[t!]
\includegraphics[width=.9\columnwidth]{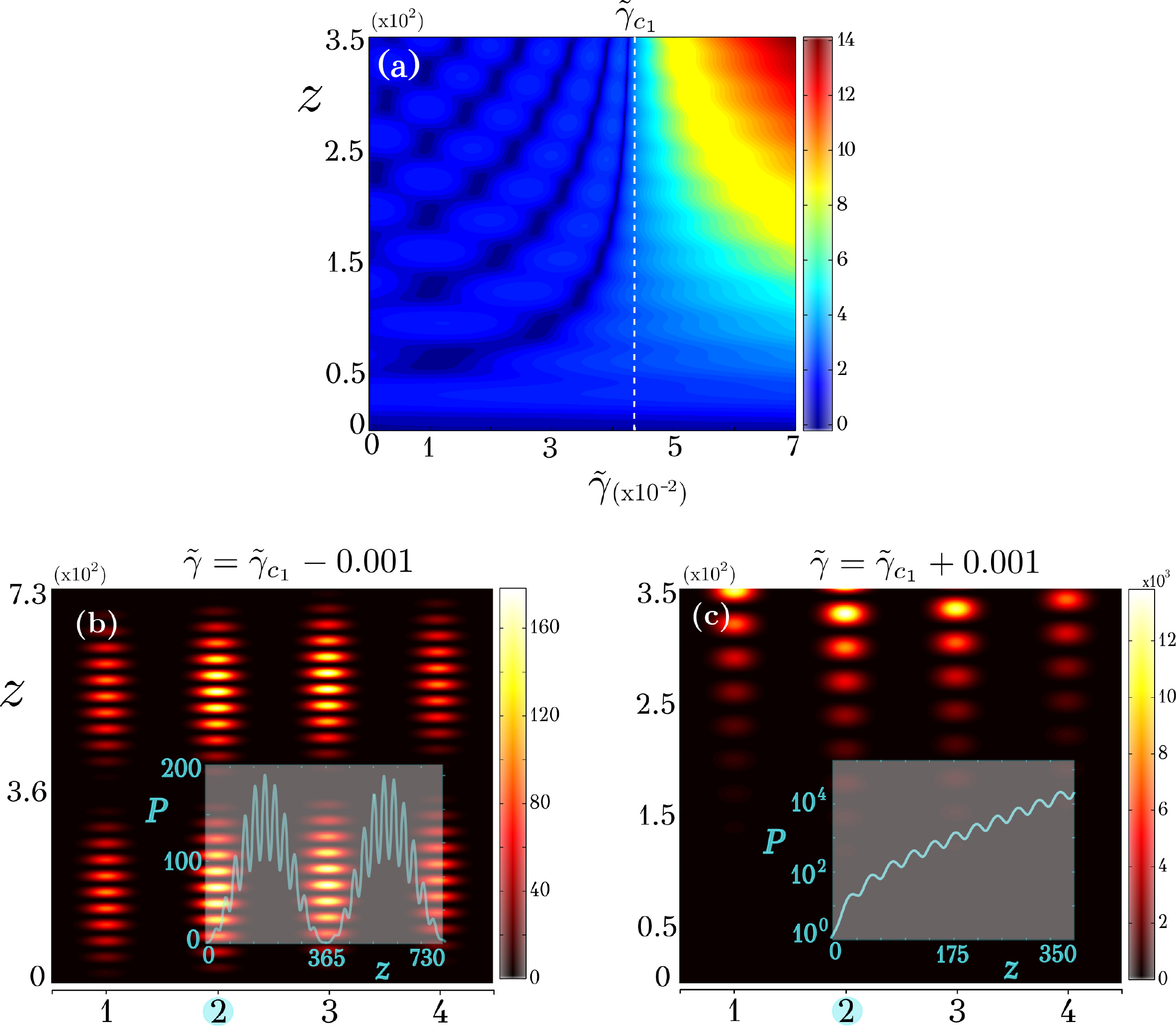}
\caption{\label{fig:EP_vicinity} (color online)  
(a) Power evolution as a function of the loss/gain parameter $\tilde{\gamma}$ across the $\mathcal{PT}$ phase transition around the exceptional point $\tilde{\gamma}_{c_{1}}$ (indicated by the dashed line), with intensity and power (insets) evolution shown for (b) $\tilde{\gamma}=\tilde{\gamma}_{c_{1}} - 0.001$ and for (c) $\tilde{\gamma}_{c_{1}}=\tilde{\gamma}_{c_{1}} + 0.001$.}  
\end{figure}

Figure~\ref{fig:EP_vicinity} (a) shows the evolution of the power $P$ along $z$ for a range of the loss/gain parameter $\tilde{\gamma}$ around the exceptional point (EP) at $\tilde{\gamma}_{c_{1}}$. 
For a certain $\tilde{\gamma}$ in region $I$, the power generally oscillates in a superposition of four different frequencies corresponding to the four absolute differences between the real energies ${\epsilon}_j$ of the spectrum. 
As the EP point is approached from below, the branches of the spectrum also approach each other and become pairwise doubly degenerate at the EP which is thus characterized by a single oscillation frequency. 
Marginally below the EP at $\tilde{\gamma}=\tilde{\gamma}_{c_{1}}-0.001$ we have two dominant frequencies producing a beating profile in the intensity along each waveguide as well as in the overall power.
Indeed, the light intensity is characterized by revivals in each waveguide and $P$ has two different oscillation scales.
This is shown in Fig.\,\ref{fig:EP_vicinity} (b) for a single waveguide excitation:
The power trajectory performs fast oscillations within a slowly oscillating envelope with minima close to zero for the specific choice of parameters. 
As the EP is approached the envelope frequency decreases and vanishes exactly at the EP where the field intensity and the power become periodic.
Above $\tilde{\gamma}_{c_{1}}$ the fast oscillation persists, but the negative imaginary parts in the spectrum add an exponential increase to intensity and power, as described above.

\end{document}